
%
%
%
\def\unredoffs{} \def\redoffs{\voffset=-.31truein\hoffset=-.59truein}
\def\speclscape{}
%
%
%
%
\newbox\leftpage \newdimen\fullhsize \newdimen\hstitle \newdimen\hsbody
\tolerance=1000\hfuzz=2pt
\catcode`\@=11 
\def\bigans{b }
\def\answ{b }

\ifx\answ\bigans\message{(This will come out unreduced.}
\magnification=1200\unredoffs\baselineskip=16pt plus 2pt minus 1pt
\hsbody=\hsize \hstitle=\hsize 
\else\message{(This will be reduced.} \let\l@r=L
\magnification=1000\baselineskip=16pt plus 2pt minus 1pt \vsize=7truein
\redoffs \hstitle=8truein\hsbody=4.75truein\fullhsize=10truein\hsize=\hsbody
\output={\ifnum\pageno=0 
  \shipout\vbox{\speclscape{\hsize\fullhsize\makeheadline}
   \hbox to \fullhsize{\hfill\pagebody\hfill}}\advancepageno
  \else
 \almostshipout{\leftline{\vbox{\pagebody\makefootline}}}\advancepageno
  \fi}
\def\almostshipout#1{\if L\l@r \count1=1 \message{[\the\count0.\the\count1]}
      \global\setbox\leftpage=#1 \global\let\l@r=R
 \else \count1=2
  \shipout\vbox{\speclscape{\hsize\fullhsize\makeheadline}
      \hbox to\fullhsize{\box\leftpage\hfil#1}}  \global\let\l@r=L\fi}
\fi
%
\newcount\yearltd\yearltd=\year\advance\yearltd by -1900

\def\Title#1#2{\nopagenumbers\abstractfont\hsize=\hstitle\rightline{#1}%
\vskip 1in\centerline{\titlefont #2}\abstractfont\vskip .5in\pageno=0}
\def\Date#1{\vfill\leftline{#1}\tenpoint\supereject\global\hsize=\hsbody%
\footline={\hss\tenrm\folio\hss}}
%

\def\draftmode{\message{ DRAFTMODE }\def\draftdate{{\rm preliminary draft:
\number\month/\number\day/\number\yearltd\ \ \hourmin}}%
\headline={\hfil\draftdate}\writelabels\baselineskip=20pt plus 2pt minus 2pt
 {\count255=\time\divide\count255 by 60 \xdef\hourmin{\number\count255}
  \multiply\count255 by-60\advance\count255 by\time
  \xdef\hourmin{\hourmin:\ifnum\count255<10 0\fi\the\count255}}}
\def\nolabels{\def\wrlabeL##1{}\def\eqlabeL##1{}\def\reflabeL##1{}}
\def\writelabels{\def\wrlabeL##1{\leavevmode\vadjust{\rlap{\smash%
{\line{{\escapechar=` \hfill\rlap{\sevenrm\hskip.03in\string##1}}}}}}}%
\def\eqlabeL##1{{\escapechar-1\rlap{\sevenrm\hskip.05in\string##1}}}%
\def\reflabeL##1{\noexpand\llap{\noexpand\sevenrm\string\string\string##1}}}
\nolabels
%
\global\newcount\secno \global\secno=0
\global\newcount\meqno \global\meqno=1
\def\newsec#1{\global\advance\secno by1\message{(\the\secno. #1)}
\global\subsecno=0\eqnres@t\noindent{\bf\the\secno. #1}
\writetoca{{\secsym} {#1}}\par\nobreak\medskip\nobreak}
\def\eqnres@t{\xdef\secsym{\the\secno.}\global\meqno=1\bigbreak\bigskip}
\def\sequentialequations{\def\eqnres@t{\bigbreak}}\xdef\secsym{}
\global\newcount\subsecno \global\subsecno=0
\def\subsec#1{\global\advance\subsecno by1\message{(\secsym\the\subsecno. #1)}
\ifnum\lastpenalty>9000\else\bigbreak\fi
\noindent{\it\secsym\the\subsecno. #1}\writetoca{\string\quad
{\secsym\the\subsecno.} {#1}}\par\nobreak\medskip\nobreak}
\def\appendix#1#2{\global\meqno=1\global\subsecno=0\xdef\secsym{\hbox{#1.}}
\bigbreak\bigskip\noindent{\bf Appendix #1. #2}\message{(#1. #2)}
\writetoca{Appendix {#1.} {#2}}\par\nobreak\medskip\nobreak}
%
%
\def\eqnn#1{\xdef #1{(\secsym\the\meqno)}\writedef{#1\leftbracket#1}%
\global\advance\meqno by1\wrlabeL#1}
\def\eqna#1{\xdef #1##1{\hbox{$(\secsym\the\meqno##1)$}}
\writedef{#1\numbersign1\leftbracket#1{\numbersign1}}%
\global\advance\meqno by1\wrlabeL{#1$\{\}$}}
\def\eqn#1#2{\xdef #1{(\secsym\the\meqno)}\writedef{#1\leftbracket#1}%
\global\advance\meqno by1$$#2\eqno#1\eqlabeL#1$$}
%
\newskip\footskip\footskip14pt plus 1pt minus 1pt 
\def\footnotefont{\ninepoint}\def\f@t#1{\footnotefont #1\@foot}
\def\f@@t{\baselineskip\footskip\bgroup\footnotefont\aftergroup\@foot\let\next}
\setbox\strutbox=\hbox{\vrule height9.5pt depth4.5pt width0pt}
\global\newcount\ftno \global\ftno=0
\def\foot{\global\advance\ftno by1\footnote{$^{\the\ftno}$}}
%
\newwrite\ftfile
\def\footend{\def\foot{\global\advance\ftno by1\chardef\wfile=\ftfile
$^{\the\ftno}$\ifnum\ftno=1\immediate\openout\ftfile=foots.tmp\fi%
\immediate\write\ftfile{\noexpand\smallskip%
\noexpand\item{f\the\ftno:\ }\pctsign}\findarg}%
\def\footatend{\vfill\eject\immediate\closeout\ftfile{\parindent=20pt
\centerline{\bf Footnotes}\nobreak\bigskip\input foots.tmp }}}
\def\footatend{}
%
%
\global\newcount\refno \global\refno=1
\newwrite\rfile
\def\ref{[\the\refno]\nref}
\def\nref#1{\xdef#1{[\the\refno]}\writedef{#1\leftbracket#1}%
\ifnum\refno=1\immediate\openout\rfile=refs.tmp\fi
\global\advance\refno by1\chardef\wfile=\rfile\immediate
\write\rfile{\noexpand\item{#1\ }\reflabeL{#1\hskip.31in}\pctsign}\findarg}
\def\findarg#1#{\begingroup\obeylines\newlinechar=`\^^M\pass@rg}
{\obeylines\gdef\pass@rg#1{\writ@line\relax #1^^M\hbox{}^^M}%
\gdef\writ@line#1^^M{\expandafter\toks0\expandafter{\striprel@x #1}%
\edef\next{\the\toks0}\ifx\next\em@rk\let\next=\endgroup\else\ifx\next\empty%
\else\immediate\write\wfile{\the\toks0}\fi\let\next=\writ@line\fi\next\relax}}
\def\striprel@x#1{} \def\em@rk{\hbox{}}
\def\lref{\begingroup\obeylines\lr@f}
\def\lr@f#1#2{\gdef#1{\ref#1{#2}}\endgroup\unskip}

\def\addref#1{\immediate\write\rfile{\noexpand\item{}#1}} 
\def\startrefs#1{\immediate\openout\rfile=refs.tmp\refno=#1}
\def\xref{\expandafter\xr@f}\def\xr@f[#1]{#1}
\def\refs#1{\count255=1[\r@fs #1{\hbox{}}]}
\def\r@fs#1{\ifx\und@fined#1\message{reflabel \string#1 is undefined.}%
\nref#1{need to supply reference \string#1.}\fi%
\vphantom{\hphantom{#1}}\edef\next{#1}\ifx\next\em@rk\def\next{}%
\else\ifx\next#1\ifodd\count255\relax\xref#1\count255=0\fi%
\else#1\count255=1\fi\let\next=\r@fs\fi\next}
%

%
\newwrite\ffile\global\newcount\figno \global\figno=1
\def\fig{fig.~\the\figno\nfig}
\def\nfig#1{\xdef#1{fig.~\the\figno}%
\writedef{#1\leftbracket fig.\noexpand~\the\figno}%
\ifnum\figno=1\immediate\openout\ffile=figs.tmp\fi\chardef\wfile=\ffile%
\immediate\write\ffile{\noexpand\medskip\noexpand\item{Fig.\ \the\figno. }
\reflabeL{#1\hskip.55in}\pctsign}\global\advance\figno by1\findarg}
\def\xfig{\expandafter\xf@g}\def\xf@g fig.\penalty\@M\ {}
\def\figs#1{figs.~\f@gs #1{\hbox{}}}
\def\f@gs#1{\edef\next{#1}\ifx\next\em@rk\def\next{}\else
\ifx\next#1\xfig #1\else#1\fi\let\next=\f@gs\fi\next}
\newwrite\lfile
{\escapechar-1\xdef\pctsign{\string\%}\xdef\leftbracket{\string\{}
\xdef\rightbracket{\string\}}\xdef\numbersign{\string\#}}

\def\writestop{\def\writestoppt{\immediate\write\lfile{\string\pageno%
\the\pageno\string\startrefs\leftbracket\the\refno\rightbracket%
\string\def\string\secsym\leftbracket\secsym\rightbracket%
\string\secno\the\secno\string\meqno\the\meqno}\immediate\closeout\lfile}}
\def\writestoppt{}\def\writedef#1{}
\def\seclab#1{\xdef #1{\the\secno}\writedef{#1\leftbracket#1}\wrlabeL{#1=#1}}
\def\subseclab#1{\xdef #1{\secsym\the\subsecno}%
\writedef{#1\leftbracket#1}\wrlabeL{#1=#1}}
\newwrite\tfile \def\writetoca#1{}
\def\leaderfill{\leaders\hbox to 1em{\hss.\hss}\hfill}
\def\writetoc{\immediate\openout\tfile=toc.tmp
   \def\writetoca##1{{\edef\next{\write\tfile{\noindent ##1
   \string\leaderfill {\noexpand\number\pageno} \par}}\next}}}

%
\catcode`\@=12 
%
\edef\tfontsize{\ifx\answ\bigans scaled\magstep3\else scaled\magstep4\fi}
\font\titlerm=cmr10 \tfontsize \font\titlerms=cmr7 \tfontsize
\font\titlermss=cmr5 \tfontsize \font\titlei=cmmi10 \tfontsize
\font\titleis=cmmi7 \tfontsize \font\titleiss=cmmi5 \tfontsize
\font\titlesy=cmsy10 \tfontsize \font\titlesys=cmsy7 \tfontsize
\font\titlesyss=cmsy5 \tfontsize \font\titleit=cmti10 \tfontsize
\skewchar\titlei='177 \skewchar\titleis='177 \skewchar\titleiss='177
\skewchar\titlesy='60 \skewchar\titlesys='60 \skewchar\titlesyss='60
\def\titlefont{\def\rm{\fam0\titlerm}
\textfont0=\titlerm \scriptfont0=\titlerms \scriptscriptfont0=\titlermss
\textfont1=\titlei \scriptfont1=\titleis \scriptscriptfont1=\titleiss
\textfont2=\titlesy \scriptfont2=\titlesys \scriptscriptfont2=\titlesyss
\textfont\itfam=\titleit \def\it{\fam\itfam\titleit}\rm}
 \ifx\answ\bigans\else scaled\magstep1\fi
\ifx\answ\bigans\def\abstractfont{\tenpoint}\else
\font\abssl=cmsl10 scaled \magstep1
\font\absrm=cmr10 scaled\magstep1 \font\absrms=cmr7 scaled\magstep1
\font\absrmss=cmr5 scaled\magstep1 \font\absi=cmmi10 scaled\magstep1
\font\absis=cmmi7 scaled\magstep1 \font\absiss=cmmi5 scaled\magstep1
\font\abssy=cmsy10 scaled\magstep1 \font\abssys=cmsy7 scaled\magstep1
\font\abssyss=cmsy5 scaled\magstep1 \font\absbf=cmbx10 scaled\magstep1
\skewchar\absi='177 \skewchar\absis='177 \skewchar\absiss='177
\skewchar\abssy='60 \skewchar\abssys='60 \skewchar\abssyss='60
\def\abstractfont{\def\rm{\fam0\absrm}
\textfont0=\absrm \scriptfont0=\absrms \scriptscriptfont0=\absrmss
\textfont1=\absi \scriptfont1=\absis \scriptscriptfont1=\absiss
\textfont2=\abssy \scriptfont2=\abssys \scriptscriptfont2=\abssyss
\textfont\itfam=\bigit \def\it{\fam\itfam\bigit}\def\footnotefont{\tenpoint}%
\textfont\slfam=\abssl \def\sl{\fam\slfam\abssl}%
\textfont\bffam=\absbf \def\bf{\fam\bffam\absbf}\rm}\fi
\def\tenpoint{\def\rm{\fam0\tenrm}
\textfont0=\tenrm \scriptfont0=\sevenrm \scriptscriptfont0=\fiverm
\textfont1=\teni  \scriptfont1=\seveni  \scriptscriptfont1=\fivei
\textfont2=\tensy \scriptfont2=\sevensy \scriptscriptfont2=\fivesy
\textfont\itfam=\tenit \def\it{\fam\itfam\tenit}\def\footnotefont{\ninepoint}%
\textfont\bffam=\tenbf \def\bf{\fam\bffam\tenbf}\def\sl{\fam\slfam\tensl}\rm}
\font\ninerm=cmr9 \font\sixrm=cmr6 \font\ninei=cmmi9 \font\sixi=cmmi6
\font\ninesy=cmsy9 \font\sixsy=cmsy6 \font\ninebf=cmbx9
\font\nineit=cmti9 \font\ninesl=cmsl9 \skewchar\ninei='177
\skewchar\sixi='177 \skewchar\ninesy='60 \skewchar\sixsy='60
\def\ninepoint{\def\rm{\fam0\ninerm}
\textfont0=\ninerm \scriptfont0=\sixrm \scriptscriptfont0=\fiverm
\textfont1=\ninei \scriptfont1=\sixi \scriptscriptfont1=\fivei
\textfont2=\ninesy \scriptfont2=\sixsy \scriptscriptfont2=\fivesy
\textfont\itfam=\ninei \def\it{\fam\itfam\nineit}\def\sl{\fam\slfam\ninesl}%
\textfont\bffam=\ninebf \def\bf{\fam\bffam\ninebf}\rm}
%
%
\def\noblackbox{\overfullrule=0pt}
\hyphenation{anom-aly anom-alies coun-ter-term coun-ter-terms}
\def\inv{^{\raise.15ex\hbox{${\scriptscriptstyle -}$}\kern-.05em 1}}

\def\Dsl{\,\raise.15ex\hbox{/}\mkern-13.5mu D} 
\def\dsl{\raise.15ex\hbox{/}\kern-.57em\partial}

\font\bigit=cmti10 scaled \magstep1
\def\lspace{\ifx\answ\bigans{}\else\qquad\fi}
\def\lbspace{\ifx\answ\bigans{}\else\hskip-.2in\fi} 
\def\boxeqn#1{\vcenter{\vbox{\hrule\hbox{\vrule\kern3pt\vbox{\kern3pt
	\hbox{${\displaystyle #1}$}\kern3pt}\kern3pt\vrule}\hrule}}}
\def\mbox#1#2{\vcenter{\hrule \hbox{\vrule height#2in
		\kern#1in \vrule} \hrule}}  
%

\def\grad#1{\,\nabla\!_{{#1}}\,}

\def\darr#1{\raise1.5ex\hbox{$\leftrightarrow$}\mkern-16.5mu #1}

\def\half{{\textstyle{1\over2}}} 
\def\roughly#1{\raise.3ex\hbox{$#1$\kern-.75em\lower1ex\hbox{$\sim$}}}

\def\tiny{\scriptscriptstyle}
\def\half{{1\over 2}}

\def\cross{\!\!\times\!\!}
\def\gdot{\!\cdot\!}
\font\greekbold=cmmib10

\def\bold#1{\setbox0=\hbox{$#1$}%
     \kern-.010em\copy0\kern-\wd0
     \kern.025em\copy0\kern-\wd0
     \kern-.020em\raise.0200em\box0 }
\def\bomega{\hbox{$\Omega\textfont1=\greekbold$}}
\def\grad{\bold{\nabla}}
\def\blambda{\hbox{$\lambda\textfont1=\greekbold$}}


\Title{IASSNS-HEP-93/20}{Twisted Line Liquids}

\centerline{Randall D. Kamien}
\smallskip\centerline{School of Natural Sciences}
\centerline{Institute for Advanced Study}\centerline{
Princeton, NJ 08540}\smallskip
\centerline{and}\smallskip
\centerline{T.~C.~Lubensky}
\smallskip\centerline{Department of Physics}\centerline{
University of Pennsylvania}\centerline{Philadelphia, PA 19104}\bigskip
\vskip .3in

We propose a model of directed lines where the average direction has
the nature of a cholesteric liquid crystal.  This model, for instance, would
describe the liquid of screw dislocations in the twist-grain-boundary
(TGB) phase of liquid crystals.
We show that the presence of lines does not alter the long wavelength
elasticity of a cholesteric and, therefore, does not stabilize
Landau-Peierls instability of the cholesteric phase.
We discuss other possible mechanisms for
stabilizing the twist-grain-boundary phase.

\Date{19 June 1993}
\noblackbox
\newsec{Introduction and Summary}

Directed line liquids have received attention recently in a variety of
interesting
physical situations.
Composed of lines which are compelled on average to have tangent vectors
parallel
to an imposed axis,  they
are central to the long-wavelength physics of high-temperature superconductors
in an external magnetic field, both
as entangled flux liquids \ref\drn{D.R.~Nelson, Phys. Rev. Lett. {\bf60}, 1973
(1988);
D.R.~Nelson and H.S. Seung, Phys. Rev. B {\bf 39}, 9153 (1989).}
and vortex glass states \ref\MPA{M.P.A.~Fisher, Phys. Rev. Lett.
{\bf 62}, 1415 (1989).}.
They are also of interest in nematic
polymers \ref\SB{J.V. ~Selinger and R.F. ~Brunisma, Phys. Rev. A {\bf 43},
2910, 2922 (1991).}, electrorheological fluids
and ferrofluids \ref\KN{R.D.~Kamien and D.R.~Nelson,
J. Stat. Phys. {\bf 71}, 23 (1993).}.
In this paper, we will investigate a model for chiral line liquids in which
tangent vectors of the lines
are constrained to rotate in a helical fashion along some fixed pitch
axis.

The entangled flux liquid in a superconductor is produced when thermal
fluctuations
cause the regular Abrikosov vortex lattice to melt.
The close analogy between the smectic-$A$ phase in liquid crystals and the
superconducting phase in metals and that between the nematic--to--smectic-$A$
and normal-metal--to--superconductor transition was pointed out some years ago
by de
Gennes \ref\dG{P.G.~de Gennes, Solid State Commun. {\bf 14}, 997 (1973).}.
This analogy leads to the theoretical expectation that a liquid crystal analog
of the
Abrikosov vortex phase should exist with dislocation lines replacing vortex
lines.
Recent experiments \ref\TBBe{J. Goodby, M.A. Waugh, S.M. Stein. R. Pindak, and
J.S.
Patel, Nature {\bf 337}, 449 (1988); J. Am. Chem. Soc. {\bf 111}, 8119 (1989);
G. Strajer, R. Pindak, M.A. Waugh, J.W. Goodby, and J.S. Patel, Phys. Rev.
Lett. {\bf
64}, 13 (1990); K.J. Ihn, J.A.N. Zasadzinski, R. Pindak, A.J. Slaney, and J.
Goodby,
Science {\bf 258}, 275 (1992).} confirm that such a phase does exist and that
its
properties are identical to those of the theoretically predicted TGB
(twist-grain-boundary) phase
\ref\LR{S.R.~Renn and T.C.~Lubensky, Phys. Rev. A {\bf 38}, 2132 (1988); {\bf
41
}, 4392
(1990).}\
consisting of periodically repeated twist-grain-boundaries, each composed of
periodically spaced screw dislocations whose axes rotate in a helical fashion
from
one boundary to the next.  The TGB phase can presumably melt to a chiral line
liquid
just as the Abrikosov phase melts to a directed line liquid.  The chiral line
liquid we
study in this paper is intended to model the melted TGB phase.

The entangled flux liquid phase in a superconductor is in fact the
normal-metal phase, but with enhanced viscosities.
Polymer nematics, on the other
hand differ from normal nematics in that their splay elastic constant $K_1$ is
infinite if all polymers extend from one end of the sample to the other
\ref\dgaaa{
P.G.~de~Gennes, in {\sl Polymer Liquid Crystals}, edited by A.~Ciferri, W.R.
Kringbaum,
and R.B.~Meyer (Academic, New York, 1982), Chap. 5; R.B.~Meyer, Chap. 6.  See
also
\KLN and \SB.}.
Non-linearities constrained by full rotational invariance lead \ref\Toner1{
J.~Toner, Phys. Rev. Lett. {\bf 68}, 1331 (1992).}
to a renormalized theory in which elastic constants take on a momentum
dependence.  In
particular, the twist and bend Frank constants diverge at small momentum while
the
splay modulus vanishes.
It is thus natural to ask whether there are differences between the chiral line
liquid and a cholesteric liquid crystal.  Our calculations show that the
long-wavelength properties of the chiral line liquid are identical to those of
a
cholesteric liquid crystal.  A cholesteric is a one-dimensional layered
structure
with a long-wavelength Landau-Peierls elasticity and associated destruction of
long-range periodic order \ref\lp{L.D.~Landau and E.M.~Lifshitz, Statistical
Physics, 3rd Ed., Part I, Chap. XIII (Pergamon Press, Oxford, 1980); A.~Caille,
C.R.  Acad.
Sci., Ser. B {\bf 274}, 891 (1972); T.C.~Lubensky, Phys. Rev. Lett. {\bf 29},
206 (1972).}.
The TGB phase is also a one dimensional-layered structure, and though its
elasticity is more
complex than that of a cholesteric
\ref\Toner{J.~Toner, Phys. Rev. B {\bf 43}, 8289 (1991); {\bf 46}, 5715
(1992).},\
its long-range order is also fluctuation destroyed.
A chiral line liquid has the same long-wavelength
Landau-Peierls elasticity as a cholesteric but with
elastic constant renormalized by the presence of dislocations.  The splay
elastic constant
$K_1$, which does not contribute to the Landau-Peierls energy, however,
diverges as the
square of an inverse wavenumber as it does in a polymer nematic
\ref\KLN{R.D.~Kamien, P.~Le~Doussal, and D.R.~Nelson, Phys. Rev. A {\bf
45}, 8727 (1992).}.

In this paper we add explicit degrees of freedom to describe the dislocation
line
configurations.  Though the energy of a smectic-$A$ screw dislocation tilted
a small angle $\alpha$ away from the director goes as
$\alpha^2\log(1/\alpha)$, we will assume that we are at sufficiently long
scales that the tipping energy becomes analytic \ref\smec{This follows, for
instance,
from Toner's theorem, J. Toner, private communication.}.  In section II we
derive
a Landau theory for a melted, twisted line liquid, generalizing work on
directed line
liquids.  In section III
we derive the effective long-wavelength
theory and find expressions for the new, effective Frank constants of the
cholesteric.  Finally
in section IV we comment on a possible way for
the non-harmonicity of the tipping energy to
stabilize the Landau-Peierls instability of the twist-grain-boundary phase.

\newsec{Derivation of Model}
Our system is a collection of lines described by their locations ${\bf R}_i(s)$
and parameterized
by arclength $s$ embedded in and interacting with a cholesteric liquid
crystal with a unit director field ${\bf n}({\bf r})$.
We model the lines by a local density field $\rho$ and a local tangent-density
vector
$\bf m$.
In terms of the paths of the lines ${\bf R}_i(s)$,
\eqn\erho{\rho ({\bf r}) = \sum_i\int ds\,\delta^d({\bf R}_i(s)-{\bf r}) ,}
and
\eqn\em{{\bf m}({\bf r}) = \sum_i\int ds\,{d{\bf R}_i(s)\over ds}
\delta^d({\bf R}_i(s)-{\bf r}).}
If our lines do not terminate, then $\bf m$ must be divergenceless:
\eqn\emm{\eqalign{\grad\gdot{\bf m}
&= \sum_i\int ds\, {d{\bf R}_i(s)\over ds}\gdot\grad\delta^d({\bf R}_i(s)-{\bf
r})\cr
&=-\sum_i\int ds\,{d\over ds}\delta^d({\bf R}_i(s)-{\bf r}) = 0 .\cr}}
We will not consider the case of finite lines because dislocations
cannot end within a sample.

In the TGB phase, screw dislocations align parallel on average along the local
nematic
director ${\bf n}$.  We assume this remains true in the line liquid phase and
propose
the following free energy to describe the chiral line liquid:
\eqn\efree{F= \int d^3\!r\,{A\over 2}({\bf m} - \rho_0{\bf n})^2  + F_n[{\bf
n}] ,}
where
\eqn\enem{F_n[{\bf n}] = \half\int d^3\!x\, \left[ K_1\big(\grad\gdot{\bf
n}\big)^2
+ K_2\big({\bf n}\gdot[\grad\cross
{\bf n}] - k_0\big)^2 + K_3\big({\bf n}\cross[\grad\cross{\bf n}]\big)^2\right]
.}
is the usual Frank free energy for a chiral liquid crystal
and the constraint $\grad\gdot{\bf m} = 0$ is understood  The free energy
favors ${\langle\bf m\rangle}=\rho_0{\bf n}$.  We could have included an
anisotropic tensor coupling of the form
\eqn\ean{F=\half\int d^3\!x\,(m_\mu-\rho_0n_\mu)I^{\mu\nu}(m_\nu-\rho_0n_\nu).}
By rotational invariance $I^{\mu\nu} = A\delta^{\mu\nu} + Bn_\mu n_\nu$.  If we
keep
only terms up to quadratic order in the fields, we are left only with the
isotropic
tensor $\delta^{\mu\nu}$.
In
principle, the self-coupling of the vector can be different along the pitch
axis of the twist-grain-boundary phase than perpendicular to it, or in other
words, if the
pitch points along the $x$-axis,
$I^{xx}\ne I^{yy}\equiv I^{zz}$.
We might also include interactions of the defects with themselves involving
${\bf m}$
by itself.  We believe, however, that in the long wavelength limit, this model
captures the essential physics.

The ground state configuration of the liquid
crystal has a spontaneously broken symmetry.  We choose the pitch axis to lie
along the $x$-axis and the equilibrium director,
\eqn\Dir{{\bf n}_0({\bf r})
= [0,\sin(k_0 x),\cos(k_0 x)], }
to minimize the Frank Free energy.
We introduce director
fluctuations via ${\bf n} = {\bf n}_0 + \delta{\bf n}$.  If we only work
to quadratic order in fields, then because $\bf n$ is a unit vector, it is
sufficient to
consider only $\delta{\bf n}$ with $\delta{\bf n}\gdot{\bf n}_0=0$.
Since $\langle {\bf m} \rangle$ is parallel to ${\bf n}_0$ in equilibrium, we
set
\eqn\Mn{{\bf m} = \rho{\bf n}_0 + {\bf t}.}
Fluctuations of ${\bf m}$ along
${\bf n}_0$ are described by $\rho$ and fluctuations
perpendicular to ${\bf n}_0$ described by $\bf t$. We, therefore, require ${\bf
t}
\gdot{\bf n}_0=0$.  Fluctuations in
the equilibrium average of ${\bf m}$ will point along ${\bf n}_0$.  Thus
$\langle {\bf
t}\rangle =0$, and $\langle {\bf m} \rangle = \langle \rho \rangle {\bf n}_0$.
The constraint $\grad \gdot {\bf m} =0$ is now
\eqn\Constr{{\bf n}_0 \gdot \grad \rho + \grad \gdot {\bf t } = 0 .}
and our harmonic theory becomes
\eqn\edone{F=\int d^3\!r\,\left\{{A\over 2}({\bf t} -\rho_0\delta{\bf n})^2
+ {A\over 2}\delta\rho^2\right\} + F_n[{\bf n}] ,}
subject to ${\bf n_0}\gdot\grad\rho +\grad\gdot{\bf t}=0$.  We have set
$\rho = \rho_0 + \delta\rho$. and
\eqn\efhar{
\eqalign{&F_{\delta n}[\delta{\bf n}]= F_n[{\bf n}_0+\delta{\bf n}]\cr
&\;\;=\half\int d^3\!r\,\left[ K_1\big(\grad\gdot\delta{\bf n}\big)^2
+ K_2\big({\bf n}_0\gdot[\grad\cross\delta{\bf n}]\big)^2
+ K_3\big(k_0\delta{\bf n}\cross
{\bf n}_0
+ {\bf n}_0\cross\left[\grad\cross\delta{\bf n}\right]\big)^2\right] . \cr}}
We note that if $k_0=0$ then
${\bf n}_0 =\hat z$ and this theory reduces to the model derived
for directed
line liquids in \KLN .

In order to study the fluctuations in this system we,
seek a parameterization of $\bf m$ and $\rho$ satisfying the constraint
\Constr\
and the constraint ${\bf n}_0 \gdot {\bf t} = 0$.
The first constraint is satisfied by setting
\eqna\parem{$$\eqalignno{
\rho &= \rho_0-\rho_0\grad\gdot{\bf u}&\parem a\cr
{\bf t} &= \rho_0({\bf n}_0\gdot\grad){\bf u} -\rho_0({\bf u}\gdot\grad) {\bf
n}_0 ,
&\parem b\cr}$$}
where ${\bf u}$ is any vector.
This equation is invariant under the transformation
${\bf u} \to {\bf u}'={\bf u }+ \grad\cross\blambda$.
This
comes about because the constraint \Constr\ only involves the longitudinal
components of $\bf t$.
There is, therefore, a gauge-like symmetry of the transverse components.
The constraint ${\bf t}\gdot{\bf n}_0=0$ implies
\eqn\ConstrC{{\bf n}_0\gdot\grad({\bf n}_0\gdot{\bf u}) =0 .}
We can solve this
constraint on $\bf u$ by choosing ${\bf n}_0\gdot{\bf u}=0$.
This is tantamount to a choice of gauge in
the sense that if we are given a $\bf u$ with ${\bf n}_0\gdot{\bf u}\ne 0$
we can, without
making $\bf u$ trivial,
choose a $\blambda$ such that ${\bf n}_0\gdot({\bf u} +
\grad\cross\blambda)=0$.
Note that any vector ${\bf u} ( {\bf r } )$ that depends only on
$x$, the component of ${\bf r}$ along the pitch axis, satisfies the constraint
\ConstrC.
Thus the free energy expressed in term of ${\bf u}$ will depend only on
gradients of
${\bf u}$ in the $yz$-plane.

\newsec{Rotational Symmetries and Long-wavelength Elasticity}

The free energy for a chiral line liquid, like that for a cholesteric, is
invariant with
respect to uniform translations and rotations.  It is precisely these
invariances that lead
to the Landau-Peierls elastic free energy and the destruction of long-range
order in a
one-dimensionally modulated structure. In order to obtain the
effective long-wavelength free energy,
it is useful to understand how uniform translations and rotations are
described by the variables ${\bf u}$ and $\delta {\bf n}$ describing deviations
from the
equilibrium ground state. Following the analysis in \ref\rtl{T.C.~Lubensky,
T.~Tokihiro, and
S.R.~Renn, Phys. Rev. A {\bf 43}, 5449 (1991)},
we write our fields $\bf u$ and $\delta{\bf n}$ as
\eqna\fields{$$\eqalignno{
{\bf u} &= \left[ \phi, \theta\cos(k_0 x), -\theta\sin(k_0 x)\right] ,&\fields
a\cr
\delta{\bf n}&= \left[f,g\cos(k_0 x), -g\sin(k_0 x)\right] ,&\fields b\cr}$$}
\noindent thus enforcing both $\delta{\bf n}\cdot{\bf n}_0= 0$ and
${\bf u}\cdot{\bf n}_0=0$.  Because
the ground state is not translationally invariant, it is convenient to
decompose
our fields into a sum over Brillouin zones.  We write for any field $X$,
\eqn\brill{X({\bf r}) = \sum_n X_n({\bf r}) e^{ik_0 nx} ,}
where $X_n({\bf r})$ only has fourier modes with $x$-components
$k_x\in(-{k_0\over
2},{k_0\over 2}]$.

We now consider
a rigid rotation described by the vector \bomega .  Under this rotation
${\bf n}_0$ becomes:
\eqn\no{\eqalign{{\bf n}_0' = &{\bf n}_0
+\left[\Omega_y\cos(k_0 x) -\Omega_z\sin(k_0 x), -
\Omega_x\cos(k_0 x),\Omega_x\sin(k_0 x)\right]\cr
&\quad-\left(\Omega_yz-\Omega_zy\right)
\left[0,k_0\cos(k_0 x),-k_0\sin(k_0 x)\right] .\cr}}
We can now compare \fields\ and \no\ to determine what fields $f$ and $g$ are
generated
by uniform rigid rotations.  We find
\eqna\si{$$\eqalignno{f&=\Omega_y\cos(k_0 x) - \Omega_z\sin(k_0 x)&\si a\cr
g&=-\Omega_x - k_0\left(\Omega_yz-\Omega_zy\right) .&\si b\cr}$$}
If we write ${\bf t}$ in
terms of $\phi$ and $\theta$ and match the same rigid rotation, we find
\eqna\sii{$$\eqalignno{\Omega_y&=\partial_z\phi&\sii a\cr
-\Omega_z&=\partial_y\phi&\sii b\cr
-\Omega_x - k_0 ( \Omega_y z - \Omega_z y )&=
-k_0 \phi + \sin(k_0 x)\partial_y\theta + \cos(k_0x)\partial_z\theta
&\sii c\cr}$$}
Then writing \si{}\ and \sii{}\ in terms of
the decomposition \brill\ gives (with $2X_{\tiny +} = X_{\tiny 1}+ X_{\tiny
-1}$
and $2X_{\tiny -} = -i(X_{\tiny 1} - X_{\tiny -1})$)
\eqna\siii{$$\eqalignno{
f_{\tiny +}&=\Omega_y&\siii a\cr
f_{\tiny -}&=\Omega_z&\siii b\cr
g_0&=-\Omega_x - k_0\left(\Omega_yz-\Omega_zy\right)&\siii c\cr
\partial_y\theta_{\tiny -} -\partial_z\theta_{\tiny +} + k_0 \phi_0 &=\Omega_x
+ k_0 ( \Omega_y z - \Omega_z y)&\siii d\cr
\partial_y\phi_0&=-\Omega_z&\siii e\cr
\partial_z\phi_0&=\Omega_y&\siii f\cr
}$$}
\noindent The free energy cannot depend on the uniform rotation angle ${\bf
\Omega}$.  It will,
however, depend on linear combinations of the fields $\phi$, $\theta$, $f$, and
$g$ that do
not depend on ${\bf \Omega}$.  We can construct such combinations with the aid
of \sii{} .
Note that ${\bf \Omega}$ only depends on $\theta$ via the combination
$\partial_y \theta_- - \partial_z \theta_+$.  The combination $\partial_y
\theta_- +
\partial_z \theta_+$ does not generate a rotation, and the free energy can
depend on it.
Finally, we note that ${\bf t}$ will be nonzero if $\partial_x \phi_0$ is
nonzero.
Thus, we expect on quite general grounds that $F$ will have the form
\eqn\ethefree{
F=\half\int d^3\!x\,\left\{\eqalign{&
\quad a\left[(\partial_yg_0-2k_0f_{\tiny -})^2 + (\partial_zg_0+2k_0f_{\tiny
+})^2\right]\cr
&\;\;+b\left[(\partial_y\phi_0+2f_{\tiny -})^2+(\partial_z\phi_0-2f_{\tiny
+})^2\right]\cr
&\;\;\;\;+c(k_0\phi_0+ \partial_y \theta_{\tiny -} - \partial_z \theta_{\tiny
+} + g_0)^2 \cr
&\;\;\;\;\;\;+ d(\partial_y\theta_{\tiny +} +\partial_z\theta_{\tiny -}
+\partial_x \phi_0 )^2 \cr
&\;\;\;\;\;\;\;\;
+e\left[(\partial_y\theta_{\tiny +})^2 + (\partial_z\theta_{\tiny +})^2
+(\partial_y\theta_{\tiny -})^2 + (\partial_z\theta_{\tiny +})^2\right]\cr
}\right\} .}
This form is dictated by the underlying rotational invariance. In terms of
the original constants in \efree\ and \enem , $a=(K_1+K_3)/2$, and
$2b=c=d=2e=A\rho_0^2$.
The values of $b,c,d$ and $e$ are related by the underlying rotational
invariance
discussed in section II.
There are other
terms as well, in addition to higher powers and derivatives of the invariant
combinations. We also note that
because $\rho=-\rho_0\grad\cdot{\bf u}$, $\phi_n$ for $n\ne 0$ will be massive.
 Likewise,
because of the energy of a splay configuration
$f_n$ will be massive for $n\ne 0$.

If we only consider the cholesteric, then $\rho_0=0$.  In this case we are only
left
with the term proportional to $a$ in \ethefree .  By integrating out
$f_{\tiny\pm}$ we
see that $\partial_zg$ and $\partial_yg$ cannot appear quadratically.  Thus
they will first appear quartically, and we find the classic Landau-Peierls
instability.
Including the lines, we see that $\theta$ appears without $x$-derivatives and
always
in combination with other fields.  Upon integrating out $\theta$ both the
terms proportional to $c$ and $d$ disappear.

The preceding analysis showed that we must keep all the modes
in the first Brillouin zone as well as $f_{\tiny +}$ and $f_{\tiny -}$.
Returning to
our proposed model \edone , we integrate out the modes in the second zone, as
well as
$\phi_0$.  We find, to leading order in all derivatives, an effective
free energy for $g_0$:
\eqn\elp{
F_g = \int d^3\!x\,\left\{\left({A\rho_0^2\over 3k_0^2}+K_2\right)(\partial_x
g_0)^2 +
{K_2+3K_3\over 8k_0^2}\left[(\partial_y^2+\partial_z^2)g_0\right]^2 \right\} .}
Thus the fluctuation $g_0$ suffers from the Landau-Peierls instability.  The
penetration depth of the smectic, $\lambda$, is
\eqn\epene{\lambda^2 = {3(K_2+3K_3)\over 8k_0^2(3K_2+A\rho_0^2)} ,}
and thus the effect of the defects is to decrease the penetration depth.
Though there is no long range order, $\lambda$ sets the scale over which
it is possible to see the dephasing of the ground state.  Consider two
defects in the same plane of constant phase.  Their mutual repulsion forces
them apart, and as they move they drag the plane with them.  Thus the effect of
the repulsion should be to make the system ``less'' ordered.

Additionally, the free energy for the remaining director mode $f_0$ is
\eqn\eremain{\half\int d^3\!x\, \left[ K_1(\partial_x f_0)^2 + {K_2+K_3\over 2}
\left[(\partial_y f_0)^2 + (\partial_z f_0)^2\right]
+\left(A\rho_0^2 + K_3k_0^2\right)f_0^2 \right] ,}
and so the $f_0$ fluctuations become massive.  In analogy with results on
polymer nematics
\KLN\ we could interpret part of this as a defect density dependent
divergent shift in $K_1$, namely
\eqn\esplay{
K_1'({\bf q}) = K_1 + {A\rho_0^2\over q_x^2} .}

\newsec{Stabilization of the Twist-Grain-Boundary Phase}

Because the Landau-Peierls instability in the twist-grain-boundary phase is
due to the underlying rotational invariance, it is hard to imagine how any
long-wavelength description of the phase could have long range order.  As
mentioned above, the energy of a defect tilted
at an angle $\alpha$  with respect to the layer normal is
proportional to $\alpha^2\log(1/\alpha)$.  While it is true that upon
integrating out
short-distance modes, the effective energy will become analytic
in $\alpha$, the coarse-graining procedure will be cutoff by length scales
inherent to the system.
In particular, as we coarse-grain, the distance between the
screw dislocations decreases.  When the distance between them becomes
on the order of the penetration depth, we must stop.  If the screw dislocations
are
close enough ({\sl i.e.} $k_0$ is sufficiently large), then the defect energy
will remain
non-analytic.  The argument in the previous section would break down, because
now we could have a term such as $-\log(q){\bf t}({\bf q})\gdot{\bf t}(-{\bf
q})$ in the free
energy in addition to those in \efree .  Though again, we would expect that
${\bf t}$
would always appear in combination with $\delta{\bf n}$, it is hard to see how
non-analytic momentum dependence would change things.  Since the energy cost
of a bend in the direction of the cholesteric pitch will pick up this
logarithmic
energy in the twist-grain-boundary phase, we might consider, as a toy model,
a slightly more rigid cholesteric described by a phase field $u$ with
free energy
\eqn\efunny{F=\half\int {d^3\!q\over (2\pi)^3}
\,u(-{\bf q})\left[B\log\vert 1/aq_x\vert q_x^2u + K(q_y^2+q_z^2)^2\right]
u({\bf q}) .}
In this model, there is still a Landau-Peierls instability, in that
\eqn\eunst{\eqalign{
\langle\,u^2(x)\,\rangle
&=\int {d^3\!q\over (2\pi)^3}\,{1\over -B\log\vert aq_x\vert q_x^2 +
K(q_y^2+q_z^2)^2}\cr
&=\int {dq_xq_\perp dq_\perp\over (2\pi)^2}\,
{1\over -B\log\vert aq_x\vert q_x^2 + Kq_\perp^4}\cr
&={1\over 32\pi\sqrt{KB}} \int_{1/L}^{1/a} {dq_x\over q_x\sqrt{-\log aq_x}}\cr
&={\sqrt{\log(L/a)}\over 16\pi\sqrt{KB}}\cr}}
and so there will still be a divergence with system size, but it will scale as
the square-root
of the logarithm instead of the logarithm itself.

Unless there are sufficiently long-ranged interactions it appears that the
underlying rotational invariance will always destabilize the
twist-grain-boundary
phase.

\newsec{Acknowledgements}

It is a pleasure to acknowledge stimulating discussions with J. Toner.
RDK was supported in part by the National Science Foundation, through Grant
No.~PHY92--45317, and through the Ambrose Monell Foundation.
TCL was supported in part by the National Science Foundation through grants
No.~DMR91-20668 and No.~DMR91-22645.

\footatend\vfill\supereject\immediate\closeout\rfile\writestoppt
\baselineskip=14pt\centerline{{\bf References}}\bigskip{\frenchspacing%
\parindent=20pt\escapechar=` \input refs.tmp\vfill\eject}\nonfrenchspacing

\bye